\documentstyle[12pt]{article}
\begin{document}
\newcommand{\rt}{\mbox{$\tilde{R}$}}
\newcommand{\rtt}{\mbox{$\tilde{\tilde{R}}$}}
\newcommand{\kb}{\mbox{$\bar{K}$}}
\newcommand{\lb}{\mbox{$\bar{\Lambda}$}}
\newcommand{\rb}{\mbox{$\bar{R}$}}
\newcommand{\cb}{\mbox{$\frac{\lb}{\kb}$}}
\newcommand{\pb}{\mbox{$\frac{\kb}{\lb}$}}
\begin{flushright}
ITP--SB--93--81
\end{flushright}

 \begin{center}

 \vskip3em
 {\bf \huge On Projective Gravity\\ and the vanishing of the \\ Cosmological
Constant}

 \vskip3em
{\large  Charidimos Gasparakis}

 \vskip1em

 {Institute for Theoretical Physics, SUNY at Stony Brook \\
 Stony Brook, NY 11794-3800 USA \\
 E-mail: gasparakis@max.physics.sunysb.edu}

\vskip1em
October 20, 1993.
 \end{center}

 \vskip 2cm

\begin{abstract}
 We generalize Einstein's Lagrangian in a non-polynomial (in R) way. The usual
Lagrangian (linear in R) is the zero $\alpha'$ limit of our theory, where
$\alpha'$ is a parameter that is  interpreted as the inverse cosmological
costant before the Planck time.  The theory space of this lagrangian admits a
${\bf Z_{2}}$ modular group, namely $R \leftrightarrow 1/R$. Independence of
the modular invariant expectation values from the number of `Big Bangs'
enforces a quantization condition for the cosmological constant. At the
semiclassical approximation we obtain $\Lambda =0$, and a vacuum equation which
is equivalent to  inflation cosmology.  D=4 and D=1 universes are obtained as
unique (and topologically separated by the D=2 semiclassical barrier) integer
dimension solutions. They  correspond to the first excited level and the ground
state respectively of our projective gravity.
 \end{abstract}
 \pagebreak

\section{Introduction}
One of the long-standing problems of physics is the quantization of gravity and
the vanishing of the cosmological constant. It has been claimed \cite{kn:col}
that the former would lead to the latter. In this paper we investigate the
theory space, and the actional defined on it, of a generalization of general
relativity, in a way that the corresponding actional admits a (projective)
representation of a ${\bf Z_{2}}$ symmetry, which we will call `duality'.
Topological terms can be introduced in this actional, which lead to a
quantization condition for the cosmological constant.

This kind of duality has already been discovered in string theory. It is well
known that toroidal compactifications of strings exhibit the phenomenon of
duality \cite{kn:dua}, which means that conformal field theories based on
different (connected by duality transformation) target space sigma model
backgrounds are equivalent.
 The simplest example of this is when we have a single toroidally compactified
dimension on a circle, with equivalalent  geometries being the ones that are
connected by the global ${\bf Z_{2}}$ symmetry
\begin{equation}
R \leftrightarrow 1/(\alpha'^{2}R). \label{eq:duar}
\end{equation}

 For spherical compactifications \cite{kn:gr} the situation is not clear.
Duality equivalent surfaces of a sphere do not have to be spheres. Non abelian
duality has been conjectured to be a symmetry between different conformal field
theories.

 In this paper we consider the symmetry (\ref{eq:duar}) at the actional level.
This symmetry effectively provides a cutoff for the range of the Ricci scalar,
in the sense that physically distinct values for R belong to the interval
$[-1/\alpha',1/\alpha']$. If the parameter $\alpha'$ is small, then this
restriction does not affect the low gravity phenomena that we observe in the
nature, although it probably will affect the black hole physics.

The geometric meaning of this symmetry is the following.  In the case of a
metric $g_{\mu \nu}$ of constant curvature $R$ we can perform a conformal
transformation

\begin{equation}
 \hat{g}_{\mu\nu}=(\alpha' R)^{2} g_{\mu\nu}, \label{eq:stm31}
\end{equation}
to obtain a metric of inverted curvature:

\begin{equation}
R(\hat{g})=1/(\alpha'^{2} R). \label{eq:stm32}
\end{equation}

In general, inverting the curvature by a conformal transformation

\begin{equation}
\hat{g}_{\mu \nu}=\Omega ^{2} g_{\mu \nu}, \label{eq:stm33}
\end{equation}

 is equivalent to solving the following differential equation (\cite{kn:bir})
for $\Omega (x)$ in D dimensions:

\begin{equation}
R'=\frac{1}{\alpha'^{2}R}=\Omega^{-2} R+2(D-1)\Omega^{-3} \Box \Omega
+(D-1)(D-4)\Omega ^{-4} (D \Omega)^{2}. \label{eq:stm34}
\end{equation}

To restore the original value of $R$ one has just to perform the inverse
conformal transformation $\Omega^{-2}$ on $\hat{g}$.
 It would be interesting  to see if we can implement this symmetry at the
Lagrangian level. The usual conformal gravity \cite{kn:sie} is one solution,
which however has bigger symmetry than the one we want. In this letter we
consider a theory space of Langrangians  which manifestly (and minimally) has
this symmetry. Its form is inspired from the form of the projective
transformations on a Riemann surface, and therefore the corresponding theory we
baptize `projective gravity'.

The gravitational coupling constant whose inverse squared multiplies the usual
Einstein term has in D dimensions mass dimension equal to

\[ [\kappa ]=\frac{2-D}{2}, \;\;\;\;\; K \equiv \frac{1}{\kappa^{2}}. \]

 In string theory we encounter the mass dimension -2 string tension $\alpha '$
, which leads us to consider the general action

\begin{equation}
 S= \int d^{D} x \sqrt{g(x)}  \frac{K(x) R(x) + \Lambda(x)}{\alpha ' R(x) + 1}.
 \label{eq:prgrgen}
\end{equation}

We observe that on the space of background fields $(K(x), \Lambda (x))$,
(dilaton and tachyon respectively, after perhaps some field redefinitions) the
$\sqrt{g}$ appears as a conformal scale. It will turn out that only conformal
classes of the background fields are relevant.
Also, the normalisation in the action is fixed by the unity in the denominator,
and the usual Einstein action (with a cosmological term) is naturally the small
$\alpha '$ limit of our projective action.  Higher power in $R$ actions have
been considered in the past (e.g. \cite{kn:tom}). Our action however is
non-polynomial.

Let us consider the following global GL(2,R) transformation:
\begin{equation}
 R \mapsto R'= \frac{a R + b/\alpha '}{c \alpha ' R +d}, \label{eq:prtr}
\end{equation}
performed by using a conformal transformation $\Omega^{2}$ on the metric, as in
(\ref{eq:stm34}).
The transformed action has the same form if

\begin{equation}
b+d=1,  \label{eq:con1}
\end{equation}
to preserve the scale.
The transformed under duality coefficients have the following form :

\begin{equation}
K'= (a K+c \Lambda \alpha')\Omega^{D} , \label{eq:eq1}
 \end{equation}

\begin{equation}
\Lambda '=(b \frac{K}{ \alpha '} +(1-b) \Lambda)\Omega^{D}  , \label{eq:eq2}
\end{equation}

\begin{equation}
(\alpha ')' = \alpha ' (a+c) . \label{eq:eq3}
\end{equation}

In the above formulae, the factor $\Omega^{D}$ comes from the conformal
transformation of $\sqrt{g}$, which equivalently can be thought of as acting
(as a conformal transformation again) on the space of the background fields.
By demanding that the string constant is invariant by duality, we obtain the
further restriction
\begin{equation}
a+c=1 .   \label{eq:con2}
\end{equation}

We can readily verify that the elements of GL(2,R) that satisfy
(\ref{eq:con1}), (\ref{eq:con2}), which we denote as $S(a,b)$ form a group too:

\[ \left( \begin{array}{cc}  a & b/\alpha' \\ (1-a)\alpha' & 1-b \end{array}
\right)  \left( \begin{array}{cc}  a' & b'/\alpha' \\ (1-a')\alpha' & 1-b'
\end{array} \right) = \left( \begin{array}{cc}  a'' & b''/\alpha' \\
(1-a'')\alpha' & 1-b'' \end{array} \right), \]
with
\[a''=aa'+b(1-a') \;\;\;\;\;\;\; b''=ab'+b(1-b'), \]
and also

\[ \left( \begin{array}{cc}  a & b/\alpha' \\ (1-a)\alpha' & 1-b \end{array}
\right)^{-1} =  \left( \begin{array}{cc}  \frac{1-b}{a-b} &
\frac{-b}{(a-b)\alpha'} \\ (1- \frac{1-b}{a-b})\alpha' & 1- \frac{-b}{a-b}
\end{array} \right), \]
provided that the determinant of the transformation
\[detS(a,b) \equiv det \left( \begin{array}{cc}  a & b/\alpha' \\ (1-a)\alpha'
& 1-b \end{array} \right) =a-b, \]
  is not zero.

This group acts on the space of $K, \Lambda$
\[V_{C} \equiv \left( \begin{array}{c} K \\ \Lambda \end{array} \right) \]
in the projective transpose representation:
\[V_{C}'=\Omega^{D}S^{t}(a,b)V_{C}.\]

Let us mention at this point that if we further restrict our attention to the
subgroup that satisfies
 \begin{equation}
S^{t}(a,b)=S^{-1}(a,b), \label{eq:frb}
\end{equation}

 we are left with $\bf{ Z}_{2}$, spanned by $S(1,0)=\bf{1}$ and $S(0,1) \equiv
S_{d}$.

We now concentrate on the $\alpha' R' =1/(\alpha'R)$ case, which corresponds to
the following transformation matrix:

\begin{equation}
 S_{d}= \left( \begin{array}{ll}  0 & 1/\alpha' \\ \alpha' & 0 \end{array}
\right), \;\;\;\; S_{d}^{2}=1.  \label{eq:duatrama}
\end{equation}

Then by applying the above formulae (\ref{eq:eq1}),(\ref{eq:eq2}) we obtain
\begin{equation}
\Lambda '=\frac{K}{\alpha'}\Omega^{D} ,  \label{eq:duatrama21}
\end{equation}

\begin{equation}
K'=\Lambda \alpha ' \Omega^{D}.  \label{eq:duatrama22}
\end{equation}

The conclusion is that we obtain two duality covariant conformal densities of
weight $2D$:
\begin{equation}
 \Lambda K, \;\;\;\; (\Lambda K)'=(\Lambda K)\Omega^{2D},  \label{eq:charg}
 \end{equation}

\begin{equation}
 \Lambda^{2}+(K/\alpha')^{2}, \;\;\; ( \Lambda^{2}+(K/\alpha')^{2})'= (
\Lambda^{2}+(K/\alpha')^{2})\Omega^{2D}, \label{eq:charg2}
 \end{equation}

with mass dimensions 2(D-1) and 2D respectively. What is important is that the
choice of (\ref{eq:duatrama}) is the unique one that allows the existence of
duality covariant combinations of the $\kappa ,\; \Lambda$ parameters. The
symmetry group of (\ref{eq:prgrgen}) has been reduced to $\bf{ Z}_{2}$, the
same symmetry as the one acting on the Ricci scalar. Another way to see this is
to intoduce the quantity

\begin{equation}
P \equiv \frac{K}{\Lambda} ,\label{eq:P}
\end{equation}

which transforms as a scalar under $\bf{Z_{2}}$ (zero conformal weight):
\begin{equation}
P'=\frac{\alpha'^{2}}{P}. \label{eq:P1}
\end{equation}

Another interesting observation is that the transformation laws
(\ref{eq:duatrama21}), (\ref{eq:duatrama22})  act on the determinant that
appears in the Lagrangian (\ref{eq:prgrgen}) in a projective orientation
reversing way:

\begin{equation}
(detL)' \equiv K'-\alpha' \Lambda' =(-K+\alpha' \Lambda')\Omega^{D} =-(det
L)\Omega^{D}. \label{eq:d}
\end{equation}

\section{Actional approach}

In this section we will construct the theory space of our gravity theory
(\ref{eq:prgrgen}), which will be endowed with a fiber bundle structure, and we
will introduce an actional \cite{kn:hata} on it. Single valuedness of the
expectation values of the ${\bf Z_{2}}$ modular invariant quantities over the
(compactified) range of the Ricci scalar will provide us  a quantization
condition for the charge. This single valuedness has the physical
interpretation that the cosmological constant spectrum is independent of the
number of times that the Big Bang took place.
Let us now be explicit:

We define:

\begin{eqnarray}
V=\left( \begin{array}{c} V_{L} \\ V_{C} \end{array} \right) \\
V_{L} =\left( \begin{array}{c} V^{1} \\ V^{2} \end{array} \right) =\left(
\begin{array}{c} \rt \\ \rtt\  \end{array} \right) \\
V_{C} =\left( \begin{array}{c} V^{3} \\ V^{4} \end{array} \right)  =\left(
\begin{array}{c} K \\ \Lambda \end{array} \right) \\
m=1,2 \;\;\;\;\; a=3,4
\end{eqnarray}

with
\begin{equation}
R=\frac{\rt}{\rtt}  \;\;\;\; [R]=[\rt]=2 \;\;\;\; [\rtt]=0 ,  \;\;\;\;\;  S=
\int d^{D} x  \frac{K \rt + \Lambda \rtt}{\alpha ' \rt + \rtt}. \label{eq:prac}
\end{equation}

The geometric meaning (if any) of \rt, \rtt $\;$ is not clear to us. However in
the gauge \rtt =1, \rt $\;$ is the usual curvature scalar.

Under duality the coordinates of the bundle transform as

\begin{equation}
V'=\left( \begin{array}{cc} S(a,b) & 0 \\ 0 & \Omega^{D} S^{t}(a,b) \end{array}
\right) V.
\label{eq:prtrm}
\end{equation}

In (\ref{eq:prac}) we see that actually the base manifold of our bundle (with
coordinates $V_{L}$) is {\bf RP1}=$S^{1}$. We denote the coordinate on $S^{1}$
still as $R=|R|e^{i \theta}$.  On the fiber we introduce an inner product with
metric given by $S_{d}$:
\begin{equation}
K \Lambda =\frac{1}{2} (K,\Lambda) S_{d} \left( \begin{array}{c} K \\ \Lambda
\end{array} \right).  \label{eq:prod}
\end{equation}

This metric is to be conformally conserved : If in some conformal frame
\[ \partial_{\mu} (K \Lambda) =0 \]
then in every conformally equivalent frame
\[\partial_{\mu}^{(\Omega,2D)} (K \Lambda)' =0, \;\;\;\;
\partial_{\mu}^{(\Omega, 2D)} \equiv \Omega^{2D} \partial_{\mu} \Omega^{-2D}.
\]
The group of transformations that can act on the fiber and leave its metric
invariant is again the $\bf{Z_{2}}$:
\begin{equation}
V_{C}^{t}S_{d}V_{C}=  V_{C}^{t}T^{t}S_{d}TV_{C} \Longrightarrow T=\bf{1},S_{d}
\label{eq:invm}
\end{equation}

We obtain again the condition (\ref{eq:frb}). By
(\ref{eq:prtrm}),(\ref{eq:frb}) we realise that the geometric interpretation of
our bundle is that it is a frame bundle with base manifold $S^{1}$.
We introduce the natural metric on $S^{1}$:
\begin{equation}
d \theta=-i\frac{dR}{R}. \label{eq:mea}
\end{equation}

Under the duality transformation (\ref{eq:duar}), the above measure changes
sign, which means that the orientation is reversed. If the indegrand is a
conformal density, then we must introduce appropriate factors of $\Lambda K$ in
the measure, to make the integration operation well defined on conformal
classes.

\begin{equation}
d^{(q)} \theta \equiv -i\frac{dR}{R}\frac{1}{(\Lambda K)^{\frac{q}{2D}}},
\;\;\;\; f^{(q)} \mapsto \Omega^{q} f^{(q)} ,\;\;\;\; \delta(\int d^{(q)}\theta
f^{(q)}) =0  \label{eq:mea3}
\end{equation}

Another quantity that naturally appears in the study of projective maps is the
Schwartzian derivative, defined as
\begin{equation}
D_{R}L^{(0)} \equiv \frac{L^{(0)}'''}{L^{(0)}'}-\frac{3}{2}(
\frac{L^{(0)}''}{L^{(0)}'} )^{2}  \label{eq:sch}
\end{equation}

when it acts on a scalar of zero conformal weight, and as

\begin{equation}
 D_{R}^{(\Omega,D)} \equiv \Omega^{D} D_{R} \Omega^{-D}  \label{eq:duabd}
\end{equation}

when it acts on densities of conformal weight $D$.
The prime here denotes (functional) differentiation with respect to $R$.
The nice thing about this derivative is that its kernel is the set of
fractional linear functions.
Before we define the actional, let us introduce some convenient notation:

\begin{equation}
 \bar{R} \equiv {\alpha' R}, \;\;\; \bar{K} \equiv K/\alpha', \;\;\;
\bar{\Lambda} \equiv \Lambda ,\;\;\;\;\bar{P} \equiv \frac{P}{\alpha'}.
\label{eq:duab}
\end{equation}

\begin{equation}
 \bar{K}'=\bar{\Lambda}, \;\;\; \bar{\Lambda}'=\bar{K},\;\;\; \bar{R}' =
\frac{1}{\bar{R}},\;\;\;\; \bar{P}'=\frac{1}{\bar{P}},\;\;\;\; \
L_{pr}=\frac{\bar{K}\bar{R}+\bar{\Lambda}}{\bar{R}+1}.  \label{eq:conve1}
\end{equation}

and the $*$-duality operation on the space of possible (even off-shell)
lagrangians, which are generic functions of homogeneity degree one in the
$V_{C}$ coordinates, as

\begin{equation}
 L^{*}(\bar{R};\bar{K},\bar{\Lambda})= L^{*}(\bar{R};V_{C}) \equiv
L(\bar{R};V'_{C})=\Omega^{D}L(\bar{R};\bar{\Lambda},\bar{K}), \;\;\;\;
L^{**}=\Omega^{2D} L.  \label{eq:duaa}
\end{equation}

 In general
\begin{equation}
 L(\bar{R};\bar{K},\bar{\Lambda},)=\sum_{n} \bar{R}^{n}
a_{n}(\bar{K},\bar{\Lambda},\alpha')\; , \label{eq:conv3}
\end{equation}

where the $a_{n}$ are homogeneous functions of degree one , therefore densities
of weight $D$. We also impose the normalisation condition (in order to make the
conformal dimensions right)

\begin{equation}
 \sum_{n=-\infty}^{+\infty} a_{n}(\bar{\Lambda},\bar{K},\alpha')a_{-n}
(\bar{K},\bar{\Lambda},\alpha')=\bar{\Lambda} \bar{K} .\label{eq:conv4}
\end{equation}

Then

\begin{equation}
\Omega^{-D} L^{*}(\bar{R};\bar{K},\bar{\Lambda},\alpha')=\sum_{n} \bar{R}^{n}
a_{n}(\bar{\Lambda},\bar{K},\alpha')  \label{eq:conv3a}
\end{equation}

with the same normalisation condition.

On shell (projective lagrangians)  we have that
\begin{equation}
 \Omega^{-D}  L_{pr}^{*}(\bar{R};V_{C}) =  L_{pr}(1/\bar{R};V_{C}).
\label{eq:duac}
\end{equation}

Finally, since $L^{**}=\Omega^{2D}L$, the $*$ operator has eigenvalues $\pm
\Omega^{D}$ and a complete set of eigenvectors, generically denoted as
$L_{\sigma},L_{\alpha}$ with
\begin{equation}
 L_{\sigma}^{*}=\Omega^{D}L_{\sigma}, \;\;\;\;
L_{\alpha}^{*}=-\Omega^{D}L_{\alpha}.  \label{eq:duaca}
\end{equation}

A general lagrangian (with conformal weight $D$) can be projected down to these
eigenspaces as follows:
\begin{equation}
 L_{\sigma}=\frac{1}{2} (L+\Omega^{-D}L^{*}), \;\;\;\;
L_{\alpha}=\frac{1}{2}(L-\Omega^{-D}L^{*}).  \label{eq:duacab}
\end{equation}

We now introduce the `kinetic term' of our actional \footnote{Kinetic terms
based on the Schwartzian derivative also arrear in studies of two dimensional
quantum gravity for the prepotential f of the world sheet metric
\cite{kn:pol}}, defined on the space of lagrangians $L(V)$ on the above frame
bundle:
\begin{equation}
I_{K}(L(V))=\int _{S^{1}} C\frac{[dR]}{iR} (L^{*}(V) D_{R} L(V)+L(V)
D^{(\Omega,D)}_{R} L^{*}(V))\; . \label{eq:actact}
\end{equation}

The duality invariant conformal density $C$ will be of the form
$(\alpha')^{\beta} (\Lambda K)^{\gamma}$.
Since $[L]=D ,\; [D_{R}]=-4$, we easily see that
\begin{equation}
C=\frac{1}{\alpha' \Lambda K}\; .  \label{eq:C}
\end{equation}

Under duality

\begin{equation}
\bar{R} \mapsto \bar{R}, \;\;L \mapsto L^{*},\;\; L^{*}\mapsto L^{**} ,\;\;
D_{R} \mapsto D_{R}^{(\Omega,D)},\;\;  D_{R}^{(\Omega,D)} \mapsto
D_{R}^{(\Omega,2D)}   \label{eq:acteqf23}
\end{equation}

it is easily verified that the actional is left invariant.

The field equation obtained is
\begin{equation}
\frac{\delta I}{\delta L}=D_{R}L(V)=0 \; , \label{eq:acteq}
\end{equation}

provided that we take as functional space of variations the ones that satisfy
\begin{equation}
D_{R}\delta L(V)=0\; .  \label{eq:fs}
\end{equation}

So on-shell we obtain exactly our projective gravity action, and the constants
$K,\Lambda ,\alpha'$ appear in a natural way.

A possible topological term in the actional (which is topological in view of
the normalization condition that we will impose shortly) is
\begin{equation}
I_{(T1)}(L(V))= \int_{S^{1}} J \frac{dR}{iR} \Omega^{-D} L^{*}(R;V_{C})
L(R;V_{C}) \equiv 2\pi \langle L | L \rangle . \label{eq:topa}
\end{equation}

The conformal density $J$  can be easily calculated using dimensional analysis:

\begin{equation}
 [L^{*}]=[L]=D, \;\;\;\; J=\frac{\alpha'}{\Lambda K},  \label{eq:conv2}
\end{equation}

It is obvious that under duality, the action $I_{(T1)}$ is invariant, and also
that is well defined on conformal classes. It is also clear that the
contribution
of $I_{(T1)}$ is purely topological, in a trivial way: At the constant
curvature case where the functional integration is reduced to integration over
the zero mode, we have that

\begin{equation}
 I_{(T1)}= 2 \pi \Leftrightarrow \langle L|L \rangle =1. \label{eq:conv5}
\end{equation}

Now we look at vacuum expectation values of observables of our theory.
Our interest will be focused on expressions of the following form:
\begin{equation}
\langle exp(i \int_{S^{1}} \frac{d\rb}{i \rb} \sum_{s} \frac{c_{s}}{(\lb
\kb)^{s/2}}(L^{s}+\Omega^{-Ds}L^{*s})) \rangle. \label{eq:n1} \end{equation}
It is clear that the above expressions are duality invariant. On-shell
(projective action) they can be readily evaluated in terms of $\bar{P}$.
The s-th order term in the exponential involves integrating over
\begin{equation}
 \frac{1}{(\bar{\Lambda} \bar{K})^{s/2}} \frac{(\bar{\Lambda}
\bar{R}+\bar{K})^{s}+(\bar{K}\bar{R}+\bar{\Lambda})^{s}}{(\bar{R}+1)^{s}}
,\label{eq:topazi1}
\end{equation}
and is equal to
\begin{equation}
I_{s}=2 \pi m c_{s} (\bar{P}^{s/2}+\bar{P}^{-s/2}). \label{eq:n2}
\end{equation}
The constant m that appears above is the winding number of the $S^{1}$ base
space. We focus on the s=1 case, since it is the minimal choice. We choose
$c_{1}$=2 since under the stereographic projection the line coves the circle
two times (see fig.1). The physical meaning of m is the number of times that
the universe was created and destroyed (the number of occurences of the Big
Bang and Big Crunch). If we demand that the above physical observables to be
independent of the number that the `Creation' took place, which in mathematical
terms means invariant under $m \rightarrow m+1$, the we get the following
condition:
\begin{equation}
4 \pi i (\bar{P}^{1/2}+\bar{P}^{-1/2}) =2 \pi i n,\;\;\; n \in {\bf Z}.
\label{eq:n3}
\end{equation}

\begin{equation}
 \bar{P}^{1/2}+\frac{1}{\bar{P}^{1/2}}=n ,\;\;\;\; n \in \bf{Z}/2 ,
\label{eq:conve6}
\end{equation}

with solution

\begin{equation}
 \Lambda^{\pm}_{ n}=\frac{1}{\alpha' \kappa ^{2}} \; (\frac{2}{n
\pm\sqrt{n^{2}-4}})^{2},\;\;\;  n \in \bf{Z}/2  . \label{eq:conv6d}
\end{equation}

Note that the quantisation condition (\ref{eq:conve6}) is duality invariant
(under $\bar{P} \mapsto \frac{1}{\bar{P}}$), and that is in this sense unique.

At the semiclassical limit $n \rightarrow \infty$ we obtain that the
cosmological constant becomes either zero or infinity. The infinite result can
be considered as projective coordinate artifact. We will see shortly that the
equations of motion are inconsistent at this limit. Another limit of interest
could also be the double scaling one: $n \rightarrow \infty \; , \;\; \alpha'
\rightarrow 0 \; , \;\; \alpha' n=$constant. We will not however consider it in
this paper.

 The Euler-Lagrange equations for the action (\ref{eq:prgrgen})  (in the case
of constant background fields and constant curvature):

\begin{equation}
\frac{D}{2}(\kb \rb +\lb)-(\kb-\lb)\frac{\rb}{\rb+1}=0  . \label{eq:el}
\end{equation}

Let us consider some special cases:

If
\begin{equation}
 \lb=0 \Longrightarrow \rb=0 \;\; or \;\; \rb=\frac{2-D}{D}.  \label{eq:el2}
\end{equation}

If $\lb = \pm \infty$ we have
\[ \lb (\frac{D}{2} +\frac{\rb}{\rb +1})+\kb \rb (\frac{D}{2} -\frac{1}{\rb
+1})=0 \; , \]

which is inconsistent.

In the low quantum number region, if $\lb/\kb=1$ (n=2) then we have $\rb=-1$.

We see that (only) at the semiclassical limit (\ref{eq:el2}) we obtain the
usual Einstein equation $R=0$.

Let us now compare our model with the usual inflation scenario. In the
$\alpha'=0$ limit, our action (\ref{eq:prgrgen}) becomes

\[ \int d^{D}x \sqrt{g} (KR+\Lambda)=\int d^{D}x \sqrt{g}( \kb \rb + \lb) \; ,
\]
with equation of motion

\begin{equation}
\rb +\frac{\frac{D}{2}}{\frac{D}{2}-1}  \frac{\lb}{\kb}=0 \; . \label{eq:el3}
\end{equation}

We propose the following inflation-like cosmological model: Universe begins at
the strong-gravity quantum regime ( say $n=\pm 2, \;\; \cb= 1$) which
corresponds to a cosmological constant $\lambda \equiv \kappa^{2} \Lambda =
1/\alpha'$, which in the usual inflation model corresponds to phase  transition
at the planck scale $E_{P}$ if the $\alpha' $ parameter is of order
$E_{P}^{-2}$, as is the case for the closed string tension (fig. 2) . This
solution is to be matched with the $\rb = \frac{2-D}{D}$ solution, which if it
were thought of as coming from the usual Einstein equation, it would correspond
to an equation of the form

\begin{equation}
 (1-\frac{2}{D})\lb + \rb =0  \; .\label{eq:el4}
\end{equation}

By comparing the above equations (\ref{eq:el3}),(\ref{eq:el4}), we obtain that
this is possible only for

\begin{equation}
 \frac{\frac{D}{2}}{\frac{D}{2}-1}=  (1-\frac{2}{D}) \; ,\label{eq:el45}
\end{equation}

so therefore
\begin{equation}
n=\pm 2 \longleftrightarrow D=1  \label{eq:el5}
\end{equation}

So our model at the ground state $n=\pm 2$ is equivalent to a usual Einstein
model at D=1 dimensions, with cosmological constant $\lambda$ positive and
equal to  the inverse square of the Planck mass.

We can repeat the above analysis for a general value of initial cosmological
constant.  We denote:
\begin{equation}
c_{n}^{\pm} \equiv \frac{n \pm \sqrt{n^{2}-4}}{2}, \;\;\; \frac{\lb _{n}}{\kb
_{n}}= (c_{n}^{\pm})^{-2}, \;\;\; c_{n}^{+}=-c_{-n}^{-},\;\;\; n \in
\bf{Z}_{+}/2.   \label{eq:laso1}
\end{equation}

Then (\ref{eq:el4}) gets replaced by

\begin{equation}
(c_{n}^{\pm})^{2}(1-\frac{2}{D}) \frac{\lb}{\kb} + \rb=0.   \label{eq:laso2}
\end{equation}

This, combined with (\ref{eq:el3}) leads to

\begin{equation}
(D-2)^{2} (c_{n}^{\pm})^{2}=D^{2}.   \label{eq:laso3}
\end{equation}
or equivalently

\begin{equation}
D^{2}(1-(c_{n}^{\pm})^{-2})-4D+4=0
\end{equation}

We can restrict our attention only to the + branch, since
$c_{n}^{+}^{2}=c_{-n}^{-}^{2}$. Graphically, the solutions of this equation are
presented in fig.3.

We note that for the first excited mode of our model,

\begin{equation}
n=5/2, \;\; \lambda (t \leq t_{P})=\frac{1}{4\alpha'}, \;\; D=4.
\label{eq:laso5}
\end{equation}

We also note that if our universe was to strart its evolution at a
semiclassical regime, with $c_{+\infty}^{+}=+\infty$, this implies that $D=2$,
and that there are no other integer solutions.

\section{Conclusions}
In this paper we investigated the theory space of a gravity theory, the theory
space of which exhibits a ${\bf Z_{2}}$ symmetry. Of course this symmetry is
not a symmetry of the gravity Lagrangian itself, it is only a symmetry of the
actional. This symmetry effectively reduces the the domain of physical interest
of the Ricci scalar up to the maximum value of the inverse string tension. This
domain for R can be stereographically projected on a circle. For the low energy
world that we see this restriction is inconsequential (except perhaps for black
hole physics). We examined vacuum expectation values of duality invariant
quantities, and we discovered that the physical demand that these quantities
should be independent of the number of Big Bangs that have already occured in
the universe provides a quantization condition for them, namely a quantization
condition for the cosmological constant. The we considered the solutions of the
classical equation of motion in the vacuum. We found that they naturally
incorpor

ate the inflation scenario. By demanding consistency with the usual Einstein
gravity inflation we discovered that the D=4 universe that we live in is the
first excited mode of the third quantized theory. The ground state is a D=1
universe, and the semiclassical limit is a D=2 universe. These are the only
integer solutions.

Let us now make some further comments: Our gravity theory can easiry be coupled
to matter in the usual way:
For the scalar field $\phi$ we have the Lagrangian
\begin{eqnarray}
L(\phi;m^{2},l,k,a,b)=\sqrt{g}(g^{\mu
\nu}\partial_{\mu}\phi\partial_{\nu}\phi+m^{2}\phi^{2}+\nonumber \\
+l\phi^{3}+k\phi^{4}+aR\phi^{2}+bR\phi)
\end{eqnarray}
which is invariant under (2) provided that
\begin{eqnarray}
\phi \rightarrow \Omega^{(2-D)/2}\phi, \;\;\; m \rightarrow \Omega^{-1}m,\;\;\;
l \rightarrow \Omega^{(3D-6)/2}l, \nonumber \\
k \rightarrow \Omega^{D-4}k,\;\;\; a \rightarrow a,\;\;\; b \rightarrow
\Omega^{(2-D)/2}b
\end{eqnarray}
We observe that the transformation law on the background fields does not form a
representation of the ${\bf Z_{2}}$ group, as it was the case with the
gravitational backgrounds (20), Therefore we do not obtain any further
quantization conditions.

Similarly for the Yang Mills coupling
\begin{equation}
L(A_{\mu};g_{YM})=\sqrt{g} g^{\mu \rho}g^{\nu \sigma} tr(F_{\mu \nu}F_{\rho
\sigma})
\end{equation}
we find that
\begin{equation}
A_{\mu} \rightarrow \Omega^{(4-D)/2} A_{\mu},\;\;\; g_{YM}A_{\mu} \rightarrow
g_{YM}A_{\mu},
\end{equation}
so again we can consistently couple to a scalar field without obtaining more
quantization conditions. The above arguments can be repeated also for spinors
by noticing that when we make a scale transformation the spin connection  does
not transform, only the vierbein. We can also look at higher order curvature
terms that are needed as  higher loop counterterms, e.g.
$\sqrt{g}R^{2},\sqrt{g}R_{\mu \nu}R^{\mu \nu},\sqrt{g}R_{\mu \nu \rho
\sigma}R^{\mu \nu \rho \sigma}$ and see that they transform with a factor
$\Omega^{D-4}$, so at D=4 the do not transform at all. This might be a nice
feature since these term appear as counterterms in the usual theory.

It is also of interest to try to find analogues of our symmetry in the context
of usual Riemannian geometry. Consider the usual form of metric \cite{kn:wei},
which has the group SO(3) acting as isometries on it \cite{kn:haw}:

\begin{equation}
ds_{1}^{2}=B(r)dt^{2}-A(r)dr^{2}-r^{2}(d\theta^{2}+sin^{2}\theta \; d\phi^{2}).
\label{eq:stm}
\end{equation}

We introduce coordinates
\begin{equation}
X_{1}(R)=R(r)sin\theta\;cos\phi \;\;\;\; X_{2}(R)=R(r)sin\theta\;sin\phi
\;\;\;\;  X_{3}(R)=R(r)cos\theta .  \label{eq:harco}
\end{equation}

These coordinates (which locally approximate the manifold with a sphere) are
harmonic provided $R(r)$ satisfies

\begin{equation}
\frac{d}{dr}(r^{2}B^{1/2}A^{-1/2} \frac{dR}{dr})-2A^{1/2}B^{1/2}R=0
\label{eq:harcon}
\end{equation}

Now let us perform the following global transformation:
\begin{equation}
\tilde{R}=( \alpha')^{k-1} R^{k} , \label{eq:trr}
\end{equation}

with special case of interest $k=-1$.

Then (\ref{eq:harcon}) is left invariant provided that

\begin{equation}
\tilde{B}=\frac{B}{kR^{2k-2}} \;\;\;\;\;\; \tilde{A}=kA.  \label{eq:trfi}
\end{equation}

The conclusion is that $X_{i}(1/R)$ are harmonic coordinates for the metric (in
the impotent $k=-1$ case):

\begin{equation}
ds_{2}^{2}=-R^{4} B(r)dt^{2}+A(r)dr^{2}-r^{2}(d\theta^{2}+sin^{2}\theta \;
d\phi^{2}). \label{eq:stm2}
\end{equation}

We call this the dual metric to the original one. Note that the radial part of
space has switched roles with time.
We see that (locally) we can make diffeomorphism and conformal transformation
on the $r,t$ plane to identify their metrics, but then we would loose the
identification in the $\theta, \phi $ plane, up to a multiplicative factor.
 Namely the metric in (\ref{eq:stm2}) is conformally equivalent to

\begin{equation}
ds_{3}^{2}=- B(r)dt^{2}+A(r)dr^{2}-\tilde{r}(r,t)^{2}(d\theta^{2}+sin^{2}\theta
\; d\phi^{2}). \label{eq:stm3}
\end{equation}

Note that the notions of space and time have been interchanged. It is as if
theories with particles moving with  speed smaller than the speed of light at
small $R$ are equivalent to tachyons at large $R'=1/R$.

 Let us conclude with a speculation: In the early days of quantum gravity when
people were looking at the potential renormalisability of Einstein's action, it
had been proposed that the identity
\[\frac{1}{1-\kappa ^{2} \Box}=\sum_{n} \kappa ^{2n} \Box ^{n} \]
should somehow be used to circumvent the problem that in four dimensions
$[\kappa]=-1$. This has never been made clear, but in view of our action
(\ref{eq:prgrgen}) we see that it might be indeed relevant, and effectively
being equivalent to renormalising the $\alpha '$ parameter.

\vskip1em
I would like to thank W. Siegel for helpful discussions and the ITP at Stony
Brook for providing a stimulating environment for self development.


\begin{thebibliography}{10}
\bibitem{kn:dua} A. Giveon, E. Rabinovici, G. Veneziano, Nucl. Phys. B322
(1989) 167 \\
J. Maharana, J. H. Schwarz, CALT-68-1790 \\
A. Giveon, M. Rocek, IASSNS-HEP-91/84 \\
T. Kugo, B. Zwiebach, IASSNS-HEP-92/3 \\
 W. Siegel ITP-SB-93-2
\bibitem{kn:gr} A. Giveon, M. Rocek, hep-th/9308154 \\
F. Quevedo hep-th/9305055
\bibitem{kn:bir} N. Birrel, P. Davies `Quantum fields in curved space'
Cambridge Monographs in Mathematical Physics', 1982
\bibitem{kn:col} S. Coleman, Nucl. Phys. B310 (1988) 643
\bibitem{kn:hata} H. Hata, Hep-th/9308001
\bibitem{kn:haw} S. Hawking, G. Ellis, `The large scale structure of spacetime'
Cambridge Monographs in Mathematical Physics', 1973
\bibitem{kn:pol} A.M. Polyakov, Mod. Phys. Lett. A, Vol.2 (1987) 893
\bibitem{kn:sie} W. Siegel `Introduction to string field theory', World
Scientific, Singapore 1988.
\bibitem{kn:tom} E. Tomboulis Phys. Lett B97 (1980) 77
\bibitem{kn:wei} S. Weinberg, `Gravitation and cosmology', Wiley, New York,1972




\end{thebibliography}
\end{document}